# On the 20 canonical amino acids by a cooperative vector-addition principle based on the quasi-28-gon symmetry of the genetic code

Chi Ming Yang

(Neurochemistry and Physical Organic Chemistry Program, Nankai University, Tianjin, China 300071. E-mail: yangchm@nankai.edu.cn;   Fax: + 86 22 2350 3863)

**Abstract**      Upon the covalent-bonding hybrid of the nitrogen atoms taken as a measure for the structural regularity in nucleobases, it can be identified that the internal relation within the 20 amino acids follows a cooperative vector-in-space addition principle based on the spherical and rotational symmetry of a quasi-28-gon (quasi-icosikaioctagon), with two evolutionary axes.

## 1.   INTRODUCION

The 20 standard amino acids together with 64 tri-nucleotide codons selected in the genetic code constitute a paradigm of complexity in Nature.[1] Atomic rationals for the choice of nucleobases by Nature have recently received much attention.[2]  For the importance of stereoelectronic effect in noncovalent intermolecular interaction and biomolecular recognition,[3] the nitrogen atoms of $sp^2$ hybrid in nucleobases are often the major binding sites for protons and metal ions.[4]  Upon a systematic analysis of the structural regularity of nucleobases, we used covalent bonding hybrid of nitrogen atoms, the $sp^2$ nitrogen atom hybrids (Scheme 1a), as a determinative measure for their empirical stereoelectronic property, the resulted genetic code in UCGA succession shows an almost linear correlation between amino acid property and $sp^2$ N-numbers.[5] We suggested a possible primordial core in the genetic code on the basis of two pieces of evidence.[5]  First, accompanied with a linear correlation between the hydropathy of amino acids and the $sp^2$ nitrogen-atom numbers of nucleobases in a rearranged genetic code, codons for A, V, T, G and P form a crossed-intersection core (Fig. 1a); second, a stereostructural re-classification of the 20 amino acids lead to five distinct groups of amino acids (Scheme 1b), which precisely overlap the above five groups of codons.[5] Subsequently, a physical organic chemistry principle was employed together with geometrical approaches to analyzing a rearranged code, quasi-28-gon symmetries with two presumed evolutionary axes were revealed from the code, displayed in the Yang's model for the genetic code (Fig. 1c).[5b]  The full details of the elucidation and representation of a quasi-28-gon symmetry are given in several previous papers.[5a,5b] The purpose of this communication is to describe a newly identified vector-in-space addition principle within the 20 canonical amino acids at their atomic level, which is consistent with the coevolution postulate[6] and further shows  that the atomic contents in amino acids may play a role in the origin and evolution of the genetic code.[7]

## 2. Method and results

The 64 codons consist of 16 genetic code doublets.[8a,8b] To obtain further insight into the hidden symmetries inherent in the genetic code, a series of geometrical steps to display the code were carried out,[5,8c] that is, from a three-dimensional display of the rearranged code (Figure 1a) to a closed spherical graph (Figure 1b), showing the internal relation between each two groups of the 16 genetic code doublets and their amino acids, with every line connecting two genetic code doublets which vary by one base-letter change from one to another.   Figure 1b not only exhibits rotational symmetry and spherical feature of the genetic code system, but also graphically explains the capability of the genetic code in self-maintaining its information integrity.

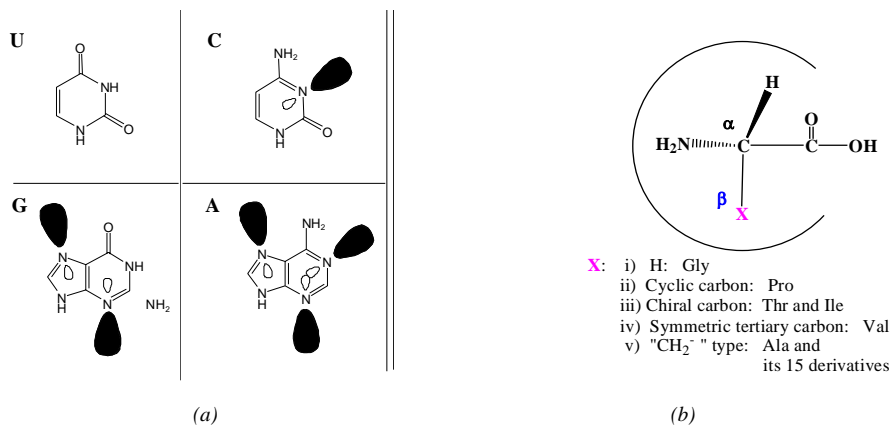

**Scheme 1**  *a*) The numbers of sp$^2$ nitrogen atoms: 3 for A, 2 for G, 1 for C and 0 for U.   *b*) Five stereostructural classes of canonical amino acids.

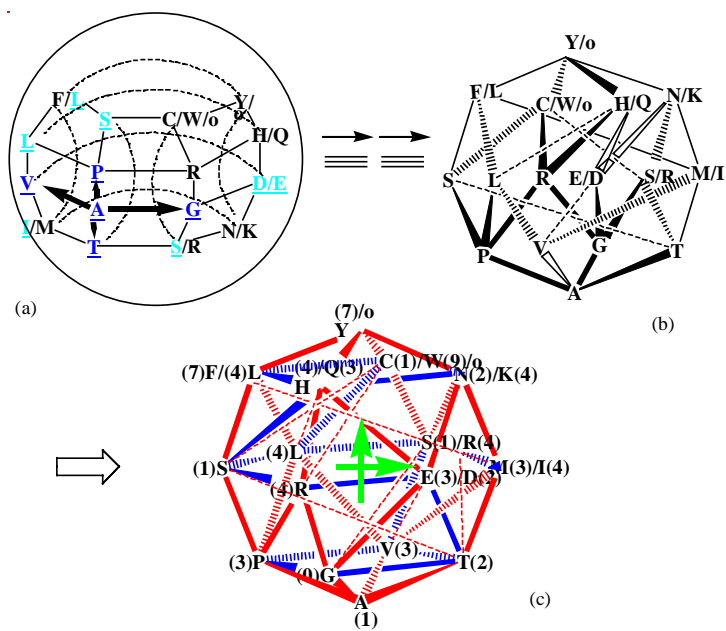

**Fig. 1** Geometrical analysis (*a* to *c*) revealed the rotational symmetries inherent in the distribution of both the number of the amino acids and their side-chain carbon-atom contents in the genetic code along a quasi-28-gon model with two presumed evolutionary axes (arrows "↑" and "→") (*c*).   In *c*, block lines (both red and blue) are edges on the polyhedron; red lines (both block and dotted) are for neighboring code-doublet connection.

Taken together both the closed spherical feature with the newly identified rotational symmetry characteristics of the code, symmetries inherent in the genetic code can be conveniently summarized in Figure 1c, by a platonic model, a quasi-28-gon.

The consequently elucidated polyhedral symmetry in the amino-acid distribution following a quasi-28-gon complies with the general even-order degeneracy constraint, which is the basic symmetry as defined for the doubly degenerate codons.  In addition to order-4 and order-6 degenerate codons in the genetic code, there are two sets of triply-degenerate codons, one of which maps onto Ile while the other maps onto "o" (stop), and two nondegenerate codons, one of which maps onto Met while the other maps onto Trp.[8a,8b]   A quasi-28-gon helps clearly indicate that slight deviations from strict symmetry have occurred at the Y/o, C/W/o and M/I genetic code doublet positions.   Despite these odd-order degenerate codons, nevertheless, the total number of amino acids at these positions remains $C_2$-symmetrical (Figure 1c). Notably, the stop codons are not totally non-sense, but allow a counterbalance for numerical distribution of both amino acids and atomic number of carbon-atoms on side-chains of amino-acid(s) along two presumed evolutionary axes, from A to Y codons and from S to I/M codons, in a quasi-28-gon model.[5]

Recently, atomic rational for the Nature's choice of nucleobases have received particular attention in theoretical study especially by Popelier and co-workers.[2] To obtain improved understanding from the quasi-28-gon symmetry of the code, here, using an analytical geometric approach to scrutinizing the internal relation within the 20 amino acids, an arithmetical regularity within these 20 amino acids is the cooperative addition-of-vectors-in-space principle within the side-chain carbon-atoms of the 20 canonical amino acids.

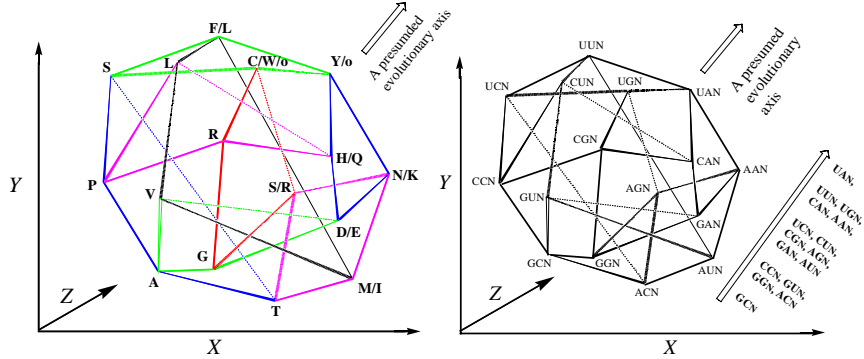

**Fig. 2** The three-dimensional feature of the code ($N$ = A, G, C and U).

Based on the polyhedral model, the three-dimensional feature of the genetic code can be summarized in Figure 2. If O = (0, 0, 0), X = (1, 0, 0); Y = (0, 1, 0), and Z = (0, 0, 1), then the vectors represented by OX($\rightarrow$), OY($\rightarrow$), and OZ($\rightarrow$) are **i**, **j**, and **k**, respectively, and are called basic vectors. Every vector in space can be written in the form $a\mathbf{i} + b\mathbf{j} + c\mathbf{k}$
in one and only one way. The numbers a, b, and c are called the first, second, and third components, respectively, of the vector.

That is, in the three-dimensional rectangular coordinate system, the unit vectors from the origin to the points (1, 0, 0), (0, 1, 0), and (0, 0, 1) are denoted, respectively, by **i**, **j**, and **k**. Any vector can be expressed in terms of these unit vectors. Thus the vector from the origin to the point P (a, b, c) is given by OP($\rightarrow$) = **A** = $a\mathbf{i} + b\mathbf{j} + c\mathbf{k}$. The vectors $a\mathbf{i}$, $b\mathbf{j}$, and $c\mathbf{k}$ are the x-, y-, and z-components of the vector **A**.

If the vectors $\mathbf{V_1}$ and $\mathbf{V_2}$ in terms of their x-, y-, and z-components are
$\mathbf{V_1} = a_1\mathbf{i} + b_1\mathbf{j} + c_1\mathbf{k}$ and $\mathbf{V_2} = a_2\mathbf{i} + b_2\mathbf{j} + c_2\mathbf{k}$
then
$\mathbf{V_1} + \mathbf{V_2} = (a_1 + a_2)\mathbf{i} + (b_1 + b_2)\mathbf{j} + (c_1 + c_2)\mathbf{k}$
and
$\mathbf{V_1} - \mathbf{V_2} = (a_1 - a_2)\mathbf{i} + (b_1 - b_2)\mathbf{j} + (c_1 - c_2)\mathbf{k}$

20 amino acids vary from one to another by their different side-chains. The numbers of side-chain carbon atoms in canonical amino acids possess very simple number 0, 1, 2, 3, 4, 7 and 9, *i.e.*, ranging from 0 for G to 9 for W. Any amino acids carrying a bigger number of side-chain carbon atoms (>2) can be the summation of two other amino acids carrying smaller numbers of side-chain carbon atoms in their ancestor codons (eq. 1).

$$aa_{N1N1N1} + aa_{N2N2N2} = aa_{N3N3N3} \qquad (1)$$

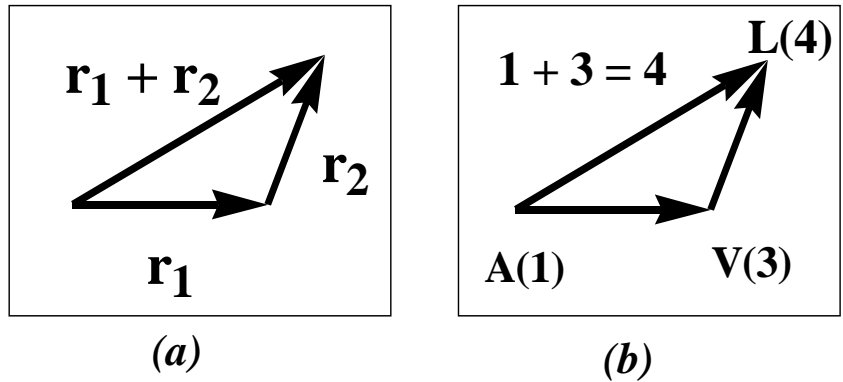

**Scheme 2** The addition-of-vectors within the amino acids: A, V and L.

According to the amino acid coding of the 16 genetic code doublets in Figure 1c, the vector-addition relation within the 20 amino acids follows one base-letter change in the 16 genetic code doublets. For example, from A codons to V codons then to L codons, there is: A(1) + V(3) = L (4), see Scheme 2.

Some obvious evidence for a vector-in-space addition relation within the amino acids at genetic code doublets along one-base-letter-change directions are depicted in Figure 3, showing:
N(2)+Y(7) = W(9);
R(4)+Q(3) = Y(7); E(3)+H(4) = Y(7); M(3)+K(4) = Y(7);
P(3)+L(4) = F(7); V(3)+L(4) = F(7); V(3)+I(4) = F(7);
P(3)+L(4) = H(4)/Q(3); P(3)+R(4) = H(4)/Q(3);
A(1)+V(3) = L(4); A(1) + P(3) = L(4);
A(1)+V(3) = I(4); A(1) + T(2) = M(3); A(1) + P(3) = R(4)

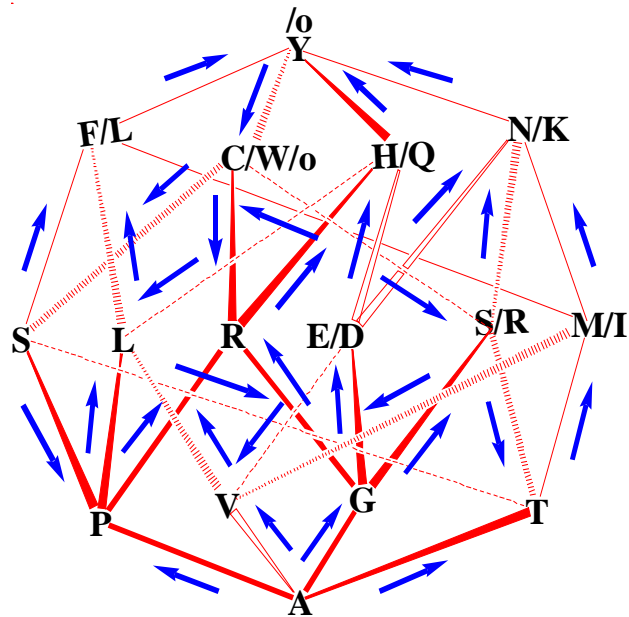

**Fig. 3** The 20 amino acids with a cooperative vector-addition principle.

As a consequence of the above observations, an improved understanding of the cooperative symmetry in the genetic code therefore is a vector-addition relation within the 20 amino acids (Figure 3). Moreover, from the directions shown by arrows, while this new finding may not reveal whether Tyr or Trp is the latest amino acid among the 20 canonical amino acids, it indicates that Ala could be the first amino acid in the codons, which is in an agreement with the recent Trivonov's proposal that Ala could be the first amino acid in the code, that has been reached previously by quite other reasoning.[9]

## 3 CONCLUSIONS

Despite the controversy over the role of physicochemical issues in shaping the code,[10] an apparently cooperative vector-addition principle at the atomic level within the 20 canonical amino acids in the genetic code is not only in conformity with the coevolution postulate in the origin of the genetic code, from simple to complex,[6] but also theoretically in agreement with the fact that every codon is formed by a tri-nucleotide, for a vector in space requires three components, *i.e.*, x-, y-, and z-components.

*(July, 2003)*